\documentclass[twocolumn]{aastex61}

\usepackage{graphicx}				
\usepackage{xcolor}
\usepackage[sort&compress]{natbib}
\usepackage[hang,flushmargin]{footmisc}
\usepackage[counterclockwise]{rotating}

\newcommand{\teff}{$T_{\rm eff}$}
\newcommand{\logg}{$\log g$}
\newcommand{\feh}{$\mathrm{[Fe/H]}$}
\newcommand{\mh}{$\mathrm{[M/H]}$}

\newcommand{\vsini}{$v \sin{i}$}
\newcommand{\tc}{$T_\mathrm{C}$}
\newcommand{\vmacro}{$v_{\mathrm{macro}}$}
\newcommand{\gcm}{g cm$^{-3}$}
\newcommand{\kms}{km s$^{-1}$}
\newcommand{\Kepler}{\textit{Kepler} }

\shortauthors{Bedell et al.}
\shorttitle{Kepler-11}

\begin{document}
\graphicspath{ {./} }
\DeclareGraphicsExtensions{.pdf,.eps,.png}

\title{Kepler-11 is a Solar Twin: Revising the Masses and Radii of Benchmark Planets Via Precise Stellar Characterization}

\author{Megan Bedell}
\affiliation{Department of Astronomy and Astrophysics, University of Chicago, 5640 S. Ellis Ave, Chicago, IL 60637, USA}

\author{Jacob L. Bean}
\affiliation{Department of Astronomy and Astrophysics, University of Chicago, 5640 S. Ellis Ave, Chicago, IL 60637, USA}

\author{Jorge Mel\'{e}ndez}
\affiliation{Departamento de Astronomia do IAG/USP, Universidade de
S\~{a}o Paulo, Rua do Mat\~{a}o 1226, Cidade Universit\'{a}ria, 05508-900 S\~{a}o Paulo, SP, Brazil}

\author{Sean M. Mills}
\affiliation{Department of Astronomy and Astrophysics, University of Chicago, 5640 S. Ellis Ave, Chicago, IL 60637, USA}

\author{Daniel C. Fabrycky}
\affiliation{Department of Astronomy and Astrophysics, University of Chicago, 5640 S. Ellis Ave, Chicago, IL 60637, USA}

\author{Fabr\'{i}cio C. Freitas}
\affiliation{Departamento de Astronomia do IAG/USP, Universidade de
S\~{a}o Paulo, Rua do Mat\~{a}o 1226, Cidade Universit\'{a}ria, 05508-900 S\~{a}o Paulo, SP, Brazil}

\author{Ivan Ram\'{i}rez}
\affiliation{McDonald Observatory and Department of Astronomy, University of Texas at Austin, USA}

\author{Martin Asplund}
\affiliation{Research School of Astronomy and Astrophysics, Australian National University, Canberra, ACT 2611, Australia}

\author{Fan Liu}
\affiliation{Research School of Astronomy and Astrophysics, Australian National University, Canberra, ACT 2611, Australia}
\affiliation{Lund Observatory, Department of Astronomy and Theoretical physics, Lund University, Box 43, SE-22100 Lund, Sweden}

\author{David Yong}
\affiliation{Research School of Astronomy and Astrophysics, Australian National University, Canberra, ACT 2611, Australia}

\correspondingauthor{Megan Bedell}
\email{E-mail: mbedell@oddjob.uchicago.edu}


\begin{abstract}

The six planets of the Kepler-11 system are the archetypal example of a population of surprisingly low-density transiting planets revealed by the \Kepler mission. We have determined the fundamental parameters and chemical composition of the Kepler-11 host star to unprecedented precision using an extremely high quality spectrum from Keck-HIRES (R$\simeq$67,000, S/N per pixel$\simeq$260 at 600 nm). Contrary to previously published results, our spectroscopic constraints indicate that Kepler-11 is a young main-sequence solar twin. The revised stellar parameters \added{and new analysis} raise the densities of the Kepler-11 planets by \replaced{about 30\%}{between 20-95\% per planet}, making them more typical of the emerging class of ``puffy'' close-in exoplanets. We obtain photospheric abundances of 22 elements and find that Kepler-11 has an abundance pattern similar to that of the Sun \added{with a slightly higher overall metallicity}. We additionally analyze the \Kepler lightcurves using a photodynamical model and discuss the tension between spectroscopic and transit/TTV-based stellar density estimates.

\end{abstract}

\section{Introduction}

Five years after their initial discovery, the six planets of the Kepler-11 system remain a crown jewel of \Kepler science results \citep[][hereafter L11]{Lissauer2011}. All six planets orbit a Sun-like host star with low eccentricies in a largely co-planar, tightly packed configuration. The formation and long-term stability of the system remains an open question \citep[see e.g.][]{Ikoma2012, Hands2014, Mahajan2014}. Kepler-11 is regarded as the prototypical example of a system of tightly-packed inner planets, a class of \Kepler multi-planet systems which offers a surprising counterpoint to our own solar system's more widely spaced architecture. Given the low geometric probability of finding a six-planet transiting system, Kepler-11 is a valuable and rare opportunity to study in detail a potentially common population of exoplanets.

In addition to their unusually tight system architecture, the Kepler-11 planets are noteworthy in another sense: their measured masses and radii place them among the lowest-density super-Earths known to date. Transit timing variations (TTVs) have been measured for all six planets. In the discovery paper,  L11 derived mass constraints for the five inner planets based on TTVs from six quarters of \Kepler data. \citet{Migaszewski2012} reanalyzed the same data using a photodynamical model and found similar results, with an additional constraint on the outermost planet's mass. The system was later revisited by \citet[][hereafter L13]{Lissauer2013} using fourteen quarters of \Kepler data. All three analyses estimate mean densities of $\leq$ 0.5 $\rho_{\earth}$ for all the planets in the system, implying a considerable gas envelope on even the smaller super-Earths. This result has implications for potential formation scenarios, with the viability of forming such low-density planets on short orbits in situ up for debate \citep[e.g.][]{Lopez2012, Chiang2013, Bodenheimer2014, Howe2015}.

Mean planet densities derived from transits and TTVs (or from transits and radial velocities) have a strong dependence on the assumed properties of the host star. Since the transit depth observationally constrains the ratio of planetary radius to stellar radius, the planet volume depends on the assumed stellar radius to the third power. The planet mass found from TTV inversion is correlated with the stellar mass. Host star characterization is therefore a critical part of measuring planet densities.

In past works, Kepler-11 has been characterized only through spectroscopic analysis of low to modest signal-to-noise data. \citet{Rowe2014}, L11, and L13 all use moderate signal-to-noise ratio spectra (S/N$\leq$ 40) from Keck and apply the Spectroscopy Made Easy package \citep[SME,][]{Valenti1996} to perform synthetic spectral fitting. The resulting stellar atmospheric parameters, when compared with stellar evolution models, indicate that Kepler-11 is a slightly evolved solar analog with a density of 0.80 $\pm$ 0.04 $\rho_{\odot}$ (L13). \added{Asteroseismic oscillation signals are not detected for Kepler-11, so the stellar density has not been independently determined \citep{Campante2014}.} Previous analysis of the stellar composition is also minimal. \citet{Adibekyan2012b} perform an equivalent width (EW) analysis on one of these Keck spectra to derive abundances of three $\alpha$-elements and find that Kepler-11 has moderately low abundances of Ca, Cr, and Ti; however, the line list employed is quite limited with $\leq$ 5 lines per element.

Kepler-11's well-characterized planetary system makes it a prime target for more detailed spectroscopic study. In this work, we present an analysis of a new, very high S/N spectrum. We use equivalent widths to measure the stellar properties and abundances of 22 elements at high precision.

The data are presented in Section \ref{s:data}. Derivation of the fundamental stellar properties from the spectrum is presented in Sections \ref{s:characterization} and \ref{s:ages}, and photospheric abundances are found in Section \ref{s:abundances}. We then present a new analysis of the \Kepler lightcurve using a photodynamical model in Section \ref{s:ttvs}. Finally, we compare the results from the spectroscopic and transit-based methods and discuss implications for the planetary system in Section \ref{s:discussion}.

\begin{table*}[ht]
\caption{Summary of derived fundamental stellar properties.}
\label{tbl:param}
\centering 
\begin{tabular}{l|cccccccc} 
\hline    
\hline 
{Spectrum}& \teff & $\sigma_{T}$ & \logg & $\sigma_{logg}$ & $v_t$ & $\sigma_{v_t}$ & \feh & $\sigma_{[Fe/H]}$ \\
{}               & (K)           & (K)                 & (dex)     & (dex)                   & (\kms) & (\kms) & (dex) & (dex)  \\
\hline
Sun (Ceres) \tablenotemark{1} & 5777 &  & 4.44 &  & 0.97 &   & 0.0 & \\
K11 & 5836 & 7 & 4.44 & 0.02 & 0.98 & 0.02 & 0.062 & 0.007 \\
HD1178 & 5650 & 7 & 4.36 & 0.02 & 0.93 & 0.02 & 0.013 & 0.008 \\
HD10145 & 5638 & 6 & 4.34 & 0.03 & 0.96 & 0.02 & $-$0.032 & 0.009 \\
HD16623 & 5791 & 26 & 4.37 & 0.07 & 0.97 & 0.06 & $-$0.462 & 0.022 \\
HD20329 & 5606 & 7 & 4.38 & 0.02 & 0.88 & 0.02 & $-$0.094 & 0.008 \\
HD21727 & 5610 & 9 & 4.38 & 0.03 & 0.96 & 0.02 & $-$0.015 & 0.007 \\
HD21774 & 5756 & 29 & 4.32 & 0.07 & 0.98 & 0.06 & 0.252 & 0.026 \\
HD28474 & 5751 & 17 & 4.47 & 0.06 & 0.93 & 0.05 & $-$0.614 & 0.014 \\
HD176733 & 5609 & 9 & 4.41 & 0.03 & 0.87 & 0.02 & $-$0.018 & 0.007 \\
HD191069 & 5710 & 7 & 4.26 & 0.02 & 1.06 & 0.01 & $-$0.044 & 0.005 \\
\hline       
\multicolumn{4}{l}{%
  \begin{minipage}{5.5cm}%
    \tablenotetext{1}{Used as reference star.}%
  \end{minipage}%
}\\
\end{tabular}
\end{table*}

\section{Data}
\label{s:data}

Owing to its relative faintness (V = 14.2, L11), previous observations of Kepler-11 were at a signal-to-noise ratio insufficient for high-precision spectroscopic characterization. We dedicated nearly 8 hours of NASA-awarded Keck I time to obtaining a higher quality spectrum. Over the course of two consecutive nights (July 26-27 2015), we made 22 1200-s exposures of Kepler-11 for a co-added result of S/N$\simeq$260 per pixel in the continuum near 600 nm. For these observations, HIRES was used with the B2 slit and kv387 filter, yielding a resolution R$\simeq$67,000 and wavelength coverage between 390 and 830 nm.

We also observed the solar spectrum (via reflection from Ceres) and nine bright potential Kepler-11 twins with the same instrumental setup and similar S/N. The Kepler-11 twins were selected by imposing criteria of 5600 $\leq$ \teff\ $\leq$ 5750 K and 4.2 $\leq$ \logg\ $\leq$ 4.4 dex on databases of previously published stellar parameters \citep{Adibekyan2012, Bensby2014}. Preference was given to stars likely to be thick-disk members with approximately solar metallicity. These criteria were set based on the original spectroscopic analysis of Kepler-11 by L11, who found \teff\ = 5680 $\pm$ 100 K, \logg\ = 4.3 $\pm$ 0.2 dex, \feh = 0.0 $\pm$ 0.1 dex, and a significant chance of Kepler-11's being a thick disk member based on its kinematics.

The spectral extraction was performed by the Mauna Kea Echelle Extraction (MAKEE) pipeline.\footnote{\url{
http://www.astro.caltech.edu/~tb/makee/}} All Kepler-11 spectra were then co-added using IRAF's \textit{scombine}.\footnote{IRAF is distributed by the National Optical Astronomy Observatory, which is operated by the Association of Universities for Research in Astronomy (AURA) under cooperative agreement with the National Science Foundation.} Continuum normalization was done by fitting low-order polynomial functions to each order, with care to use the same functional order for a given spectral order on every stellar spectrum to avoid bias in the subsequent differential analysis. Doppler corrections were applied using IRAF's \textit{dopcor} task.

\section{Stellar Properties from Spectroscopic Analysis}
\label{s:characterization}

The fundamental properties of Kepler-11 and its potential twins were derived from an equivalent width analysis. We manually measured 94 Fe I and 17 Fe II spectral lines using IRAF's \textit{splot}. The line list used unblended and unsaturated iron lines adapted from previous works such as \citet{Ramirez2014}. Laboratory values for transition probability were adopted where available, but for this strictly differential analysis the values of log \textit{gf} are largely irrelevant, since they cancel out for all lines in the linear region of the curve-of-growth. Equivalent widths were measured by carefully choosing local continua as described in \citet{Bedell2014} to maximize differential precision between the spectra. The full line list and measured equivalent widths are available in Table \ref{tbl:ews}. \added{Since our method depends on hand-measuring equivalent widths with the exact same choices made for target and reference spectra, we always use solar equivalent widths measured at the same time as the target stars. Since a few of the potential twin stars (HD10145, HD21727, and HD191069) were measured at a different time than the others, a second column of solar equivalent widths is given as a reference for those three stars only.}

The stellar effective temperature \teff, surface gravity \logg, metallicity \mh, and microturbulence $v_t$ were determined by imposing a set of requirements on the iron abundances derived by MOOG \citep{Sneden1973}. Namely, we required the \feh\ abundances from both ionization states to be equal, and any trends in iron abundance with the excitation potential or reduced equivalent width of the lines to be minimized. As the most readily observable abundant metal in the photosphere, we used iron abundance \feh\ as a direct proxy for metallicity \mh. It is important to note that we exclusively used the \textit{differential} abundance measurements relative to the solar spectrum for this analysis. By directly comparing line-by-line differential abundances of spectrally similar stars, we minimize the influence of stellar model systematics on the final parameters and abundances \citep[see e.g.][]{Ramirez2014}. 

Parameter solutions were found iteratively using the $q^2$ python package.\footnote{\url{https://github.com/astroChasqui/q2}} Uncertainties were determined by propagating scatter among the measured line abundances as described in \citet{Epstein2010} and \citet{Bensby2014}.

The resulting stellar parameters for all observed stars are given in Table \ref{tbl:param}. The \teff\ and \logg\ for Kepler-11 are significantly higher than previously determined values. We find \teff\ = 5836 $\pm$ 7 K, \logg\ = 4.44 $\pm$ 0.02 dex, and \feh = 0.062 $\pm$ 0.007 dex, while L13, for example, find \teff\ = 5666 $\pm$ 60 K, \logg\ = 4.28 $\pm$ 0.07 dex, and \feh = 0.00 $\pm$ 0.04 dex. Potential sources of this tension include the substantially different S/N of spectra used and the difference in analysis technique. L13 and other previous analyses use SME, which fits synthetic spectra to the observations. Different choices of spectral analysis technique have been shown to vary the derived stellar parameters beyond their nominal error estimates, so this explanation cannot be ruled out \citep{Hinkel2016}. However, since our analysis is performed relative to the solar spectrum, our results are anchored to the accurate stellar parameters of the Sun. Furthermore, our method is strictly differential, based on line-by-line comparison of equivalent widths measured using spectra of the Sun and Kepler-11 gathered with the same instrumentation and in the same observing run. Thus, our approach minimizes possible systematic errors that could affect other analyses.

\begin{figure}
\centering
\includegraphics[width=\columnwidth]{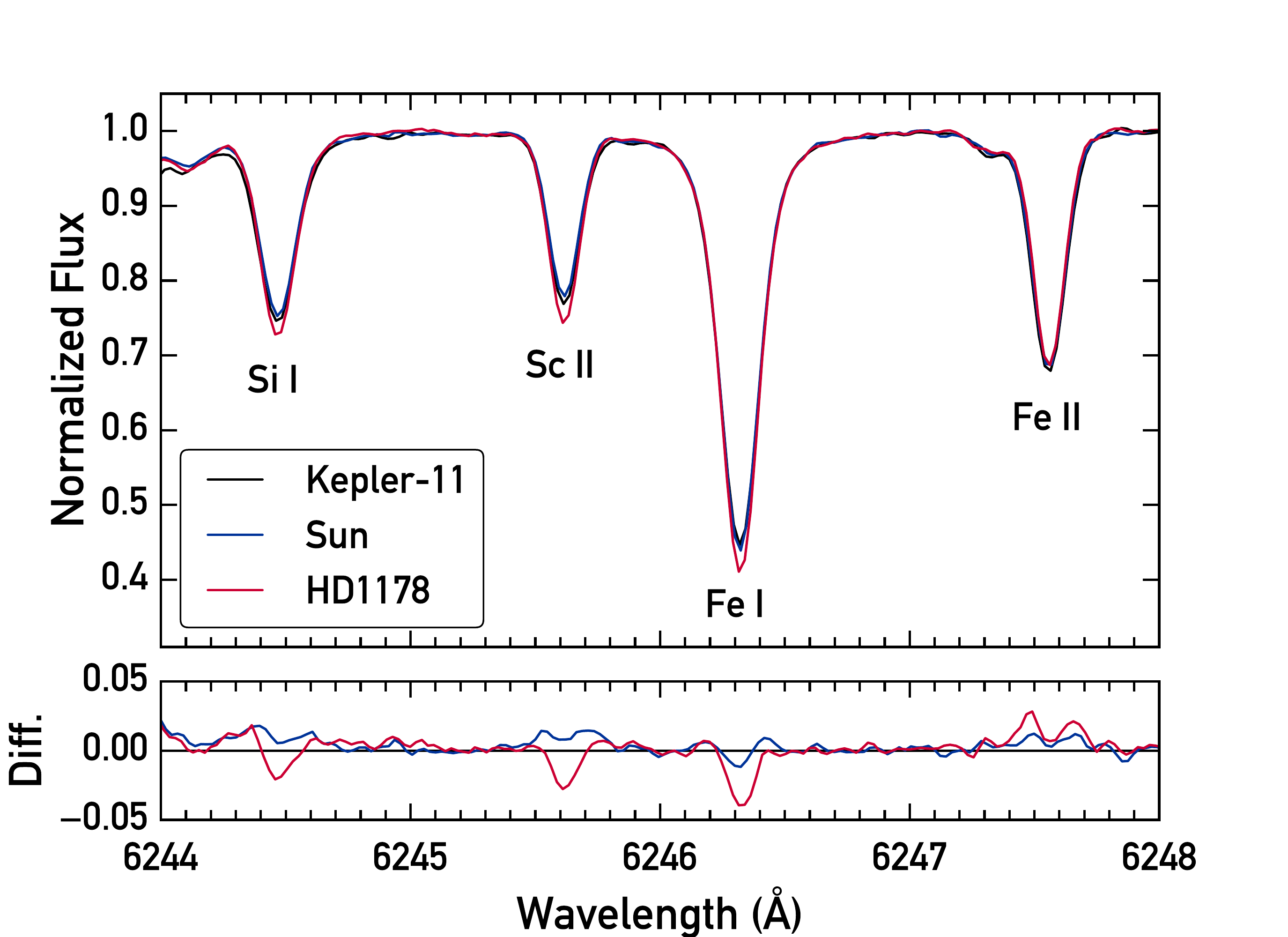}
\caption{A small section of the Keck-HIRES spectra of the Sun (blue), Kepler-11 (black), and HD1178 (red), which has fundamental parameters similar to those given by \citet{Lissauer2013} for Kepler-11. Residuals for flux relative to the Kepler-11 spectrum are plotted in the lower panel.}
\label{fig:spec}
\end{figure}

Our revised stellar parameters securely place Kepler-11 in the solar twin category. This can be seen even by eye: as depicted in Figure \ref{fig:spec}, at high S/N Kepler-11's spectrum is nearly identical to the solar spectrum and distinctly different from that of HD1178, the star from our sample whose fundamental parameters most closely match those found by L13. In particular, the solar-like \logg\ for Kepler-11 implies that it is denser and less evolved than previously thought.

\begin{figure}
\centering
\includegraphics[angle=270,width=\columnwidth]{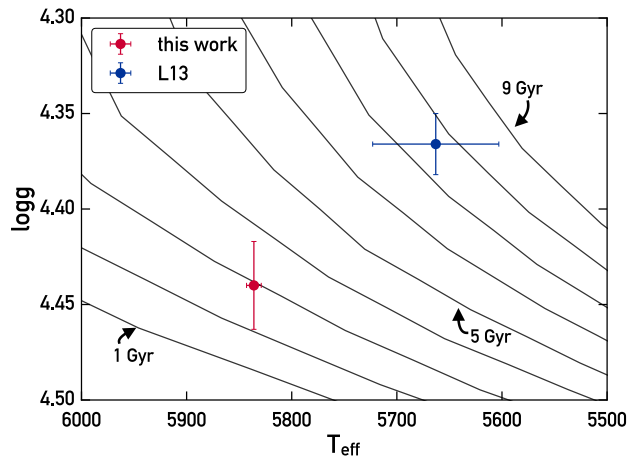}
\caption{Measured stellar properties of Kepler-11 from this work and from L13 plotted with Yonsei-Yale isochrones at a metallicity of 0.06 dex.}
\label{fig:isochrones}
\end{figure}

We used stellar evolutionary models to estimate the mass, radius, and age of Kepler-11. Yonsei-Yale isochrones were fit using $q^2$ \added{\citep{YY2004}. \replaced{We also applied Dartmouth and Basti isochrones using the \textit{isochrones} python package \citep{Morton2015}. All three models gave results consistent within 1$\sigma$. From these fits, we estimate a stellar mass $M_{\star} = 1.040 \pm 0.006 M_{\odot}$, radius $R_{\star} = 1.008 \pm 0.024 R_{\odot}$, and age $3.2 \pm 0.8$ Gyr. This gives a stellar density $\rho_{\star} = 1.43 \pm 0.10$ \gcm, or 1.01 $\pm$ 0.07 $\rho_{\odot}$.}{We applied an \feh\ offset of 0.04 dex to align the isochrone grid with the solar values, as discussed in \citet{Melendez2012}}. From this, we estimate a stellar mass $M_{\star} = 1.042 \pm 0.005 M_{\odot}$, radius $R_{\star} = 1.021 \pm 0.025 R_{\odot}$, and age $3.2 \pm 0.9$ Gyr (Figure \ref{fig:isochrones}). This gives a stellar density $\rho_{\star} = 1.38 \pm 0.10$ \gcm, or 0.98 $\pm$ 0.07 $\rho_{\odot}$.}

\added{Given the difficulty of inferring a stellar mass and radius to percent-level precision, it is worth looking deeper into the above-quoted estimates. As a solar twin, Kepler-11 is located close to the anchor point of solar-calibrated isochrone grids, so such high precision is not unreasonable. As one test, we applied Dartmouth isochrones using the \textit{isochrones} python package \citep{Dotter2008, Morton2015}. The Dartmouth models give $M_{\star} = 1.036 \pm 0.006 M_{\odot}$, $R_{\star} = 1.015 \pm 0.022 R_{\odot}$, age $3.4 \pm 0.8$ Gyr, and density 0.99 $\pm$ 0.07 $\rho_{\odot}$, in excellent agreement with the Yonsei-Yale values. While systematic errors in mass and radius may be introduced from effects like differing helium abundance and/or age-dependent gravitational settling, the proximity of Kepler-11's stellar parameters, abundances, and age to the solar values should minimize these effects. Additionally, the Yonsei-Yale grid accounts for changes in helium abundances with a metallicity-dependent scaling factor.}

\section{Alternative Stellar Age Indicators}
\label{s:ages}

\deleted{While mass, radius, and density cannot be measured through methods other than the stellar spectrum, stellar age has multiple known proxies.} \added{In addition to isochrones,} we used several alternate methods to measure the age of Kepler-11 as an independent test of its evolutionary state. The results unanimously agree upon a sub-solar age for Kepler-11. Details of the methods used follow.

\subsection{Stellar Rotation}

The apparent rotation rate \vsini\ was measured using five saturated lines (Fe I 6027.050 \r{A}, 6151.618 \r{A}, 6165.360 \r{A}, 6705.102 \r{A}, and Ni I 6767.772 \r{A}) from the Keck spectrum. The procedure used is described in depth in \citet{dosSantos2016}, and is summarized here. We first measured the macroturbulence value \vmacro$_{,\odot}$\ for each line in the solar reference spectrum using MOOG \textit{synth} with \vsini$_{\odot}$\ fixed at 1.9 \kms. We then calculated \vmacro\ for Kepler-11 using the measured solar values and an empirical relation given in Equation 1 of \citet{dosSantos2016} which calculates the expected \vmacro\ difference from the Sun as a function of stellar \teff\ and \logg. This relation was derived using 10 solar twins observed at very high resolution with HARPS, so we expect the \vmacro\ relation to be accurate for the solar twin Kepler-11 as well. Finally, MOOG \textit{synth} was used to find \vsini\ for each line in Kepler-11's spectrum with \vmacro\ fixed to the calculated value.

The five lines give a consistent result of \vsini\ = 2.2 $\pm$ 0.2 \kms. Assuming alignment of the stellar spin axis with the orbital axis of its transiting planets, we can take \vsini\ as the true rotational velocity. This translates to an age of 3.4 Gyr using the law of \citet{Skumanich1972} anchored by the Sun, or 3.0 Gyr from \citet{dosSantos2016}'s updated relation.

\subsection{Lithium Abundance}
\label{s:lithium}

\begin{figure}
\centering
\includegraphics[width=\columnwidth]{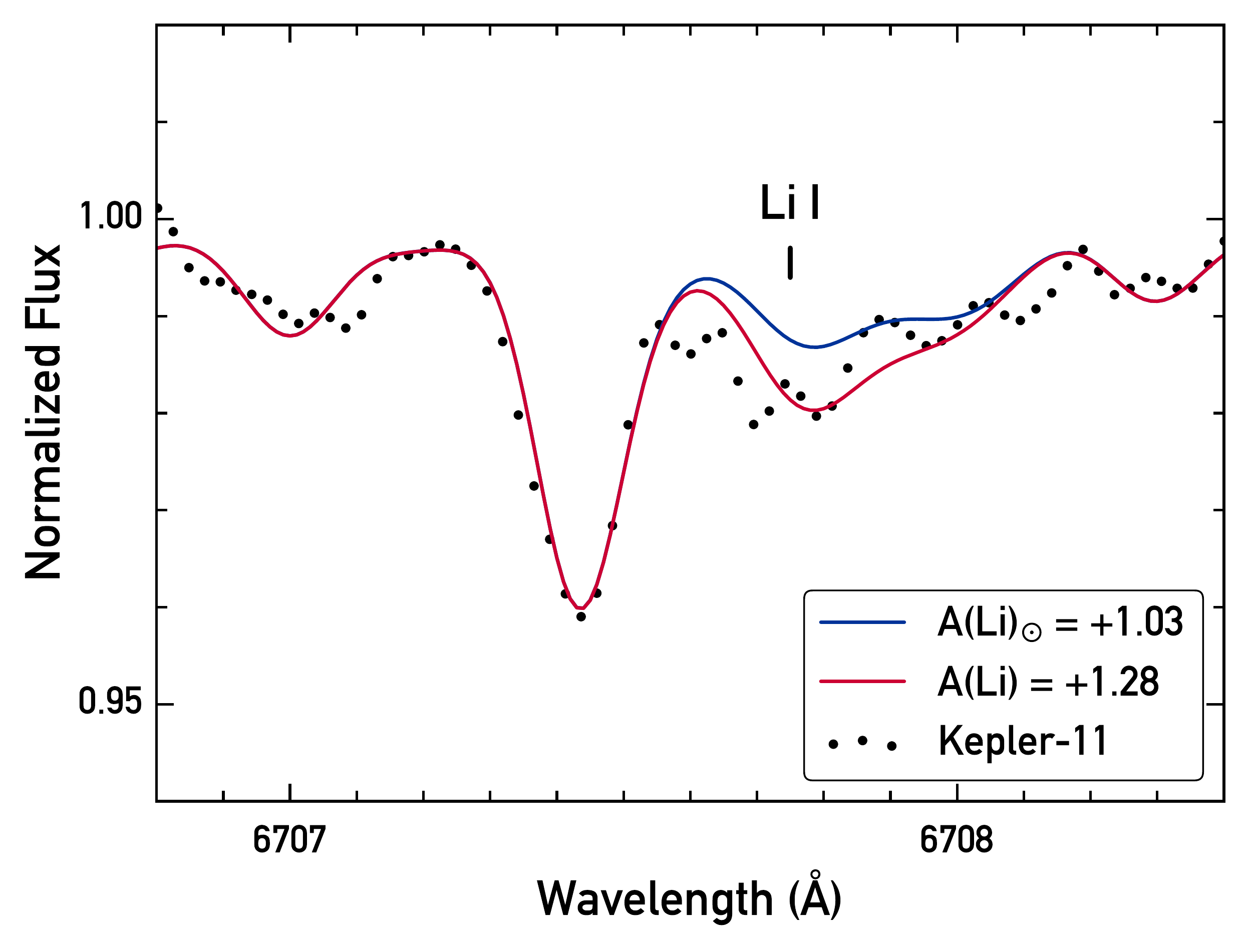}
\caption{Observed spectrum of Kepler-11 around the Li I 6707.8 \r{A} line. Synthetic fits for the best-fit Li abundance (red) and the solar Li abundance (blue) are overplotted.}
\label{fig:lithium}
\end{figure}

The lithium abundance of Kepler-11 was measured by synthesizing the Li I 6707.8 \r{A} line with MOOG \textit{synth}. The line list was adopted from \citet{Melendez2012} and includes blends of atomic and molecular lines. We find a lithium abundance of \textit{A}(Li) = 1.28 $\pm$ 0.07, higher than the measured solar value of 1.03 $\pm$ 0.04 at the level of 3$\sigma$ (Figure \ref{fig:lithium}). After applying NLTE corrections, these values become \textit{A}(Li) = 1.32 $\pm$ 0.07 for Kepler-11 and \textit{A}(Li)$_{\odot}$ = 1.07 $\pm$ 0.04 for the Sun \citep{Lind2009}.\footnote{Data obtained from the INSPECT database, version 1.0 (\url{http://www.inspect-stars.com})} Kepler-11's higher lithium abundance implies a sub-solar age, since lithium is depleted throughout a star's main-sequence lifetime \citep{Duncan1981}. Using the solar-twin-based lithium-age relation from \citet{Carlos2016} gives an age estimate of about 3.5 $\pm$ 1.0 Gyr for Kepler-11.

\vspace{8mm}

\subsection{[Y/Mg] Abundance Ratio}

Recent works by \citet{Nissen2015} and \citet{TucciMaia2016} have identified the ratio of yttrium to magnesium abundances as an excellent proxy for age in main-sequence Sun-like stars. We measured these abundances as described in Section \ref{s:abundances} and found a [Y/Mg] ratio of 0.04 $\pm$ 0.05 dex. Using the age relation from \citet{TucciMaia2016}, this gives an age of 4.0 $\pm$ 0.7 Gyr.

\subsection{Chromospheric Emission}

We measured the chromospheric emission level of Kepler-11 using the Ca II H line. Since our spectral coverage cut off around 390 nm at the blue end, it was not possible to obtain a measurement of the standard chromospheric activity index $\log(\mathrm{R'_{HK}})$. Instead, we defined an alternative index \textit{H} as the flux integrated from a 1.3 \r{A} width triangular filter centered on the H line at 3968.47 \r{A}, divided by the continuum integrated with a flat filter of 5 \r{A} width around 3979.8 \r{A}. This measurement of \textit{H} was converted to the standard Mount Wilson $S_{HK}$ using the following equation, which was derived from the literature values of ten Sun-like stars:

\begin{equation}
S_{HK} = 0.901 H + 0.033
\end{equation}

We find an activity index $\log(\mathrm{R'_{HK}})$ = -4.82. This is slightly higher than the maximum activity level of the solar cycle and suggests a sub-solar age \citep{Skumanich1972}. The activity-age relation for solar twins given in \citet{Freitas2016} yields an age estimate of 1.7 Gyr, although this is quite uncertain since we have measured the activity level at only one epoch and cannot average over the activity cycle.

\begin{figure*}
\centering
\includegraphics[scale=0.7]{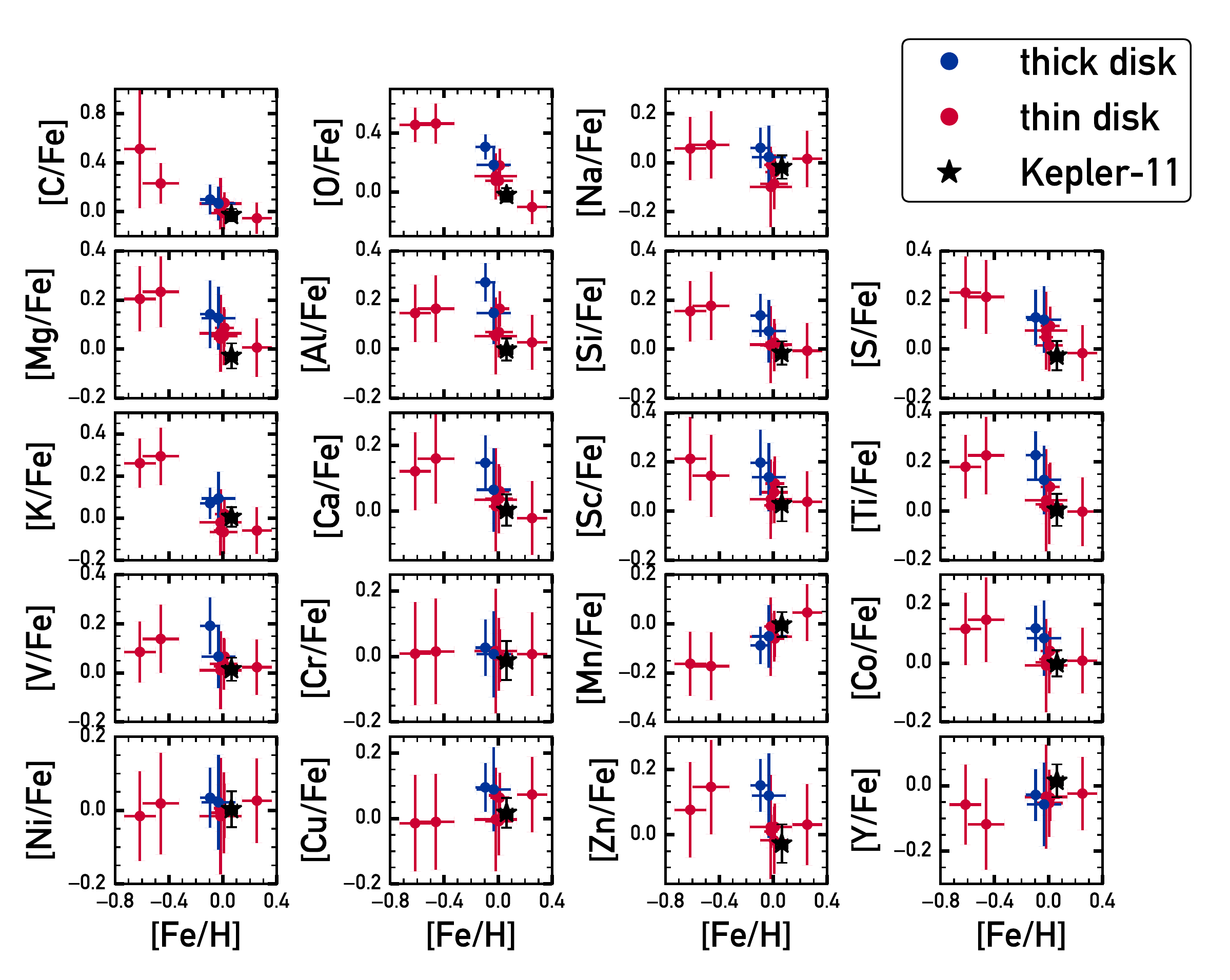}
\caption{Measured abundances plotted as a function of metallicity for the full sample. Thin (red circle) and thick (blue circle) disk stars are categorized by their kinematic membership probabilities. Kepler-11 is represented as a black star.}
\label{fig:xh}
\end{figure*}

\section{Stellar Abundances}
\label{s:abundances}

We measured photospheric abundances using the curve-of-growth technique for 20 other elements (excluding lithium, whose synthesis-based abundance determination is discussed in Section \ref{s:lithium}). As with the iron lines, all equivalent widths were measured by hand and line-by-line differential abundances determined with MOOG using $q^2$. The line list was adapted from previous works including \citet{Bedell2014}. For the element K, only one line was available, so it was measured multiple times and the deviation of the results was used as an error estimate; however, this uncertainty may be underestimated due to the line's location near a telluric-contaminated region. Hyperfine structure corrections were applied for Co I, Cu I, Mn I, V I, and Y II following \citet{Melendez2012}. Non-LTE corrections were applied for O I using grids from \replaced{\citet{Ramirez2007}}{\citet{Amarsi2015}}. Carbon abundances were measured by a combination of C I and CH lines; we note that the abundances for the two species are in tension at the $\sim$2$\sigma$ level for several of the stars in the sample, indicating that there may be some systematic effects at play. The measured equivalent widths are given in Table \ref{tbl:ews}, and resulting abundances for all stars are in Table \ref{tbl:abund}.  The quoted abundance errors include both the intrinsic scatter of the lines and the uncertainty propagated from errors on the stellar parameters. For subsequent analysis, all measured states of a given element (e.g. CI and CH, TiI and TiII, etc.) were combined with a weighted average to yield the overall elemental abundance.

\added{Since Kepler-11 was previously thought to be a potential thick-disk member based on its radial velocity (RV = -57.16 \kms\ in L11; we find -56.7 $\pm$ 0.7 \kms), several of the intended Kepler-11 twins were selected by thick-disk kinematics. As a result, we have both thin and thick disk stars in our sample. The detailed abundances of these groups can be quite different even within a small range of metallicities \citep[see e.g.][]{Liu2016}. In Figure \ref{fig:xh}, we plot the abundances for thick- and thin-disk stars as a function of their measured metallicity. Disk membership was assigned based on UVW kinematics using the procedure specified in \citet{Reddy2006}. }

\added{Kepler-11 follows the abundance trends of the other thin-disk stars well and does not display a notable $\alpha$-element enrichment. In fact, we find that despite its low radial velocity, its UVW kinematics are consistent with it being a thin disk member. Using the proper motions from UCAC3 \citep{ucac3}, our measured RV, and the isochrone-based absolute magnitude estimate ($M_V$ = 4.7 or a parallax of 1.3 mas), we find (U, V, W) = (8.1, $-$43.7, $-$6.3) \kms. This translates to a 98\% probability of Kepler-11 belonging to the thin disk population.}

Kepler-11's status as a thin-disk solar twin enables direct comparison of its abundance pattern to that of the Sun and other known solar twins. Of particular interest is the question of trends in elemental abundances with condensation temperature (\tc). As shown by \citet{Melendez2009}, the solar abundance pattern is unusual in its depletion of refractory elements relative to volatiles. This depletion has been interpreted as ``missing'' rocky material that is locked up in the Solar System planets \citep{Chambers2010}. Building up the number of stars with precisely characterized abundance patterns and planetary systems can help to test this possibility.

\begin{figure*}
\centering
\includegraphics[scale=0.6]{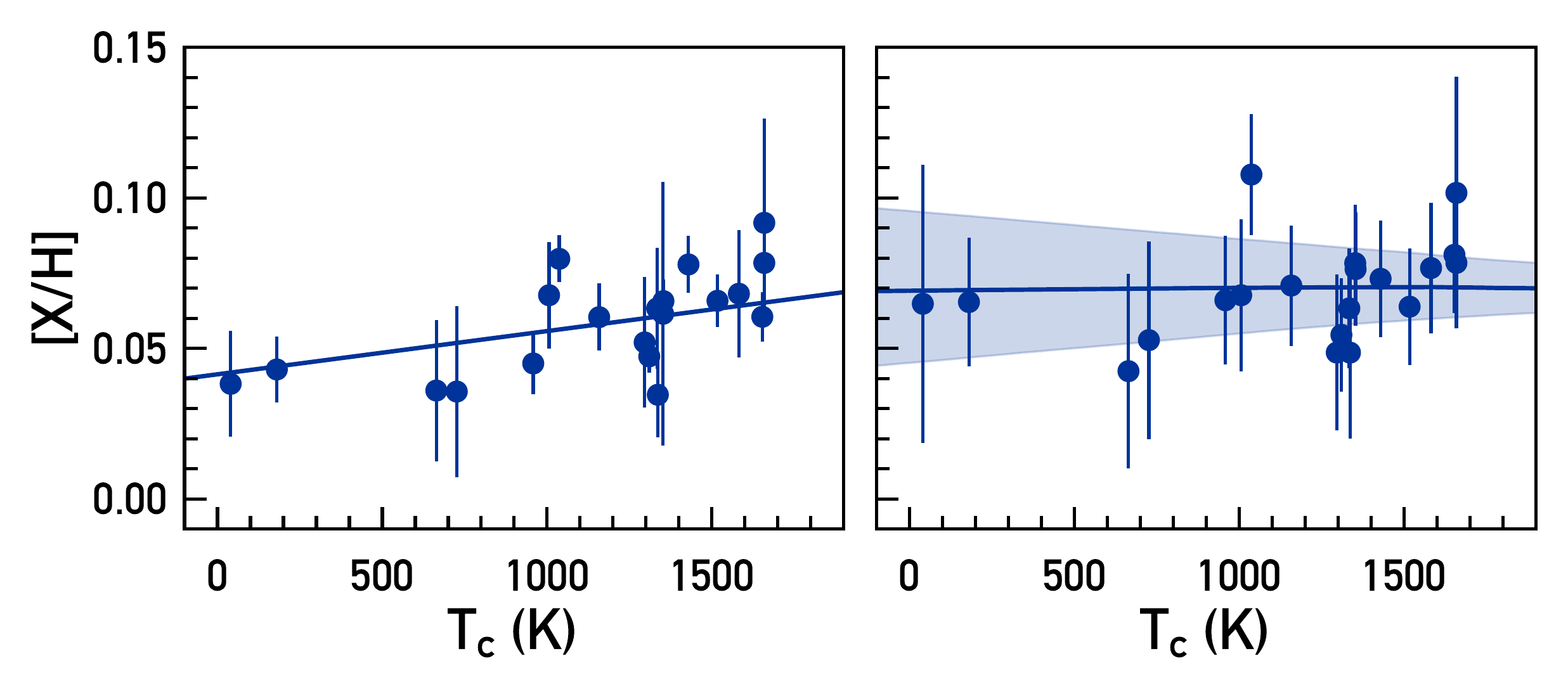}
\caption{Abundances of Kepler-11 relative to the Sun as a function of the condensation temperature of the element within the protoplanetary disk. \deleted{Error bars on the abundances come from the line-to-line scatter only (that is, not including systematic errors from the stellar parameters).} Left panel shows the abundances and best-fit linear trend before applying galactic chemical evolution (GCE) corrections, \added{with error bars from the line-to-line scatter and propagated uncertainty in the stellar parameters}. In the right panel, the data have been GCE-corrected following \citet{Spina2016b}, assuming a stellar age of \replaced{2.7}{3.2} Gyr, with error bars \replaced{from the line-to-line scatter and propagated uncertainty in the GCE correction factors}{that additionally include propagated uncertainties in the stellar age and GCE correction factors}. The shaded region represents the 1$\sigma$ uncertainty interval on the linear fit to [X/H] vs \tc\ from the bootstrap simulation described in the text.}
\label{fig:tc}
\end{figure*}

We applied corrections for the effects of galactic chemical evolution (GCE), which can change the abundance patterns and \tc\ trends of stars at varying ages \citep{Nissen2015, Spina2016}. We corrected each abundance [X/H] using the linear relationships found by \citet{Spina2016b}, who fit [X/H] as a function of stellar age for a sample of solar twins. We then used the corrected abundances and \tc\ values from Table 8 of \citet{Lodders2003} to search for a trend.

The uncertainty on the trend of [X/H] with \tc\ was propagated using a bootstrap Monte Carlo method to account for multiple potential sources of error. Each abundance is uncertain due to the intrinsic scatter of abundances derived from different lines. This uncertainty increases when the GCE correction is applied, since the correction coefficients carry some degree of random error. Additionally, the slope of the \tc\ trend can be altered by errors on the fundamental stellar parameters used \citep[as seen in][]{Teske2015} and by the uncertainty on stellar age in the GCE correction. We account for all of these effects by running 10,000 bootstrap trials where the stellar parameters are resampled from their posterior distributions; the resulting abundances are randomized by drawing samples from the multiple measured lines; the age is determined based on the resampled stellar parameters; and the GCE correction is applied using coefficients that have been randomly sampled from the (assumed Gaussian) uncertainties given in \citet{Spina2016b}. The resulting distribution of \tc\ trend fits gives a slope of [X/H] vs \tc\ of \replaced{$(-4.6^{+7.9}_{-8.7}) \times 10^{-6}$ dex K$^{-1}$}{$(-0.6^{+9.3}_{-11.0}) \times 10^{-6}$ dex K$^{-1}$} (Figure  \ref{fig:tc}). In short, the trend of Kepler-11's abundances with \tc\ is indistinguishable from the solar pattern, albeit with a large degree of uncertainty due to the many sources of error which come into play when considering GCE effects.

\added{An additional source of systematic error in the \tc\ trend is in the NLTE correction adopted for oxygen. As one of the few extremely volatile elements in our analysis (\tc\ = 180 K), oxygen has a strong influence on the \tc\ slope. The 777 nm triplet used for oxygen abundances in this analysis is also quite sensitive to NLTE effects. We carried out the above analysis using the NLTE corrections of \citet{Ramirez2007}. This yielded an oxygen abundance [O/H] = 0.058 $\pm$ 0.012 (pre-GCE correction) and a \tc\ slope of $(-4.6^{+7.9}_{-8.7}) \times 10^{-6}$ dex K$^{-1}$, still within the 1$\sigma$ uncertainty.}


\section{Stellar Properties from Photodynamic Transit Analysis}
\label{s:ttvs}

\subsection{Analysis}

In order to reassess the stellar density constraint based on the transit data, we performed a photodynamical fit to the full \Kepler short cadence (58.8 second exposure) data set. The model integrates the 7-body Newtonian equations of motions for the central star and six planets, including the light--travel--time effect. When the planets pass between the star and the line of sight, a synthetic light curve is generated \citep{2012MNRAS.420.1630P}, which can then be compared to the data. This approach therefore takes into account all transit-timing variations, simultaneously constraining planet masses, eccentricities, and radii. To prepare the data for fitting, we detrended the data with a cubic polynomial with a 2880 minute (2 day) width every 100 points, and interpolated for points between. We divided the flux by this fit as a baseline to generate our data set of 1746779 points. We additionally multiplied the uncertainties given by \Kepler by a factor of 1.115318 so that the reduced $\chi^2$ of a fiducial model was 1.0. This broadens our posteriors and helps take into account unmodeled noise in the data. To simultaneously generate the posteriors on all of our model parameters, we ran differential evolution Markov chain Monte Carlo \cite[DEMCMC, ][]{TerBraak2005} fits with planetary parameters $\{P,\, T_0,\, e^{1/2} \, \cos(\omega),\, e^{1/2} \, \sin(\omega),\, i,\, \Omega,\, R_\mathrm{p}/R_\star,\, M_\mathrm{p}/M_\star\}$ for all planets, where $P$ is the period, $T_0$ is the mid-transit time, $e$ is eccentricity, $\omega$ is the argument of periapse, $i$ is inclination, $\Omega$ is nodal angle, and $R$ and $M$ are radius and mass, respectively (with subscripts p $=$ b, c, d, e, f, g for the planets and $\star$ for the star). The star has five additional parameters: $\{M_\star,\, R_\star,\, c_1,\, c_2,\, dilute\}$, where $\{c_i\}$ are the two quadratic limb-darkening coefficients and $dilute$ is the amount of dilution from other nearby sources. We used eccentricity vector components scaling as $e^{1/2}$ so that we get flat priors in total $e$, and fixed the values of $dilute=0$ since there is no evidence of other nearby stars diluting the lightcurve. We also fixed the value of $M_\star$, as transits alone generally only give information about the density of the star, rather than $M_\star$ and $R_\star$ individually. We fixed $\Omega = 0$ for all planets because the data are not precise enough to constrain these values \citep{Migaszewski2012}. Additionally, it is extremely unlikely that there are large mutual inclinations among the planets given that we see six transiting planets (L11, Figure 4), five of which are dynamically packed and thus have no misaligned non-transiting planets between them (L11). We used flat priors for all other parameters.

We ran two DEMCMCs to model the data. One had no constraints on the stellar radius, i.e., allowed the transits themselves to completely determine the stellar density, which we will label $\mathcal{NSI}$ for ``No Spectral Information.'' The second DEMCMC was run with the stellar mass and radius fixed at the spectroscopically measured values in this study, $M_\star=1.04M_\odot$ and $R_\star=1.02M_\odot$, which we will label $\mathcal{FSP}$ for ``Fixed Stellar Parameters.'' The $\mathcal{NSI}$ run produces a lower density star $\rho_\star = 1.19^{+0.04}_{-0.11}$ g cm$^{-3}$ than the fixed value of $\rho_\star = 1.38$ g cm$^{-3}$ in $\mathcal{FSP}$. This indicates that the transit data alone are discrepant with the spectroscopically measured stellar density. Table~\ref{table:den} shows the \added{mass, radius, and} density results for all bodies for both DEMCMC runs. We note that the densities of planets with no spectral information, $\mathcal{NSI}$, are slightly higher than reported in L13 because that study includes the lower spectroscopically measured stellar density in their final best fits.

The best fit solution from $\mathcal{NSI}$ run has a lower $\chi^2$ value by more than 40 compared to the best-fit $\mathcal{FSP}$ run. Thus we see that fixing the stellar parameters at their spectroscopically measured values causes the fit to the \Kepler data to become significantly worse; the p-value for such an increase in $\chi^2$ is on order $10^{-9}$. This confirms the existence of tension between the transit measured stellar density and the spectroscopically measured one. 

\vspace{8mm}

\subsection{Physical Interpretation}

Transit measurements of stellar (and thus planet) densities rely on the the transit of the planet probing the width of the star. For a given stellar mass, once the period of a planet is known from successive transits its orbital velocity ($v_\mathrm{orb}$) can be determined. The physical distance a planet traverses during the duration of a transit ($T_\mathrm{dur}$) is to a very good approximation $T_\mathrm{dur} / v_\mathrm{orb}$. There are two main degeneracies between the stellar radius and and the measured duration: (1) eccentricity of the planets orbit and (2) impact parameter of the transit. 

Eccentricity changes $v_\mathrm{orb}$ as a function of orbital phase following Kepler's Second Law. However the observed transit timing variations provide information on the level of eccentricity of the interacting planets, and they are all found to be very small ($<0.05$), only negligibly affecting the measured stellar radius. 
Using standard orbital mechanics, it may be seen that $\rho_\star \propto R_\star^{-3} \propto v_\mathrm{orb}^{-1} = (G M_\star (\frac{2}{r} - \frac{1}{a}))^{-1/2} \propto 1 - e \sin \omega + \mathcal{O}(e^2)$, where $G$ is the Newtonian gravitational constant, $a$ is the planet's semi-major axis, and $r$ is the instantaneous star-planet distance. Thus a change in $\rho_\star$ by the $\sim$20\% required to reconcile the spectroscopic and TTV measurements would require a uniform increase in $e \sin \omega$ across all planets of order 0.06, well beyond that allowed by the TTVs.  
Our fits marginalize over the range of possible eccentricities by including the eccentricity vectors as free parameters when fitting for stellar and planetary densities. In the $\mathcal{FSP}$ DEMCMC, the planets' eccentricities do increase substantially, but the chains are unable to find a TTV solution nearly as good as for the low eccentricity case, as discussed above. 

\begin{figure}
\centering
\includegraphics[width=\columnwidth]{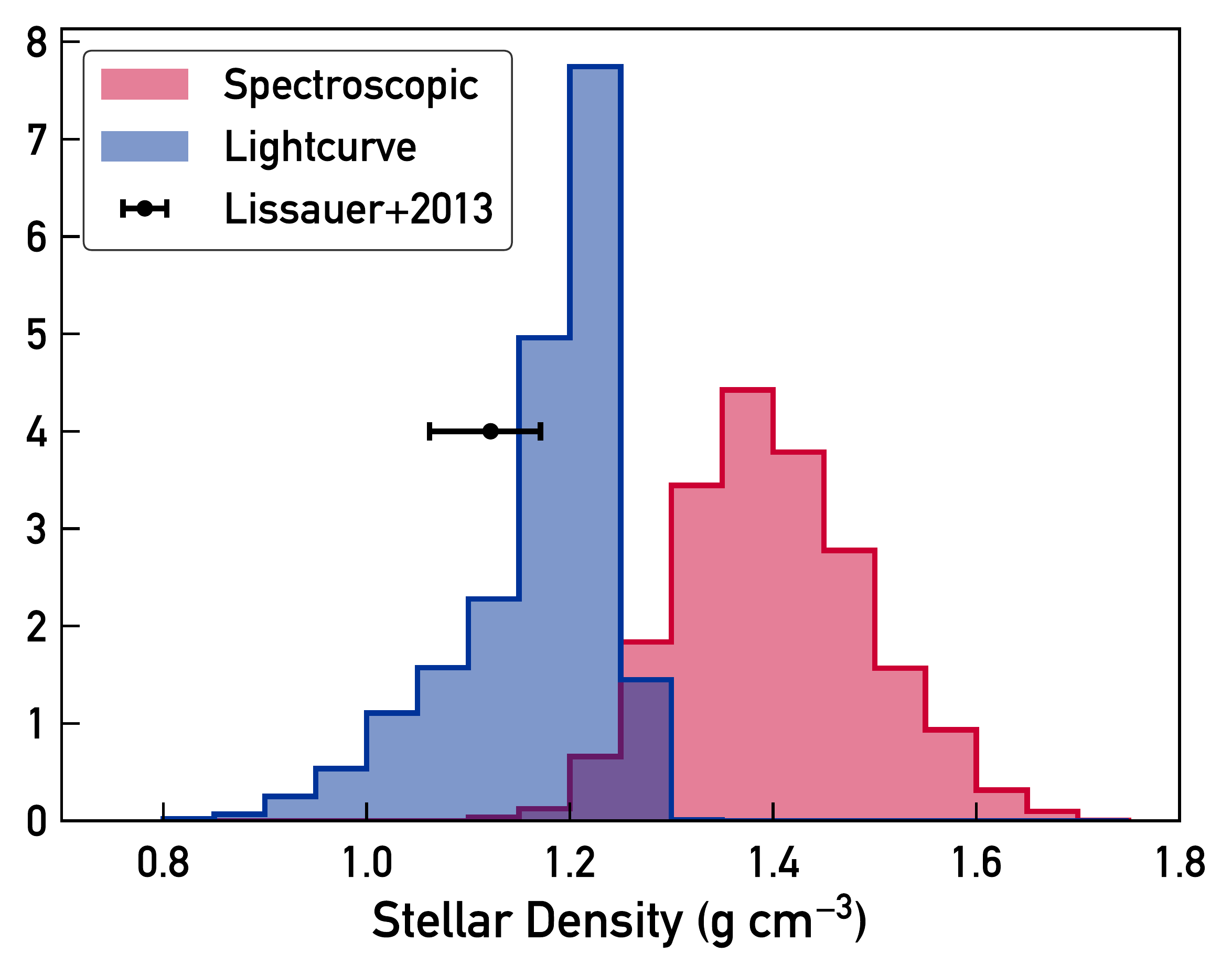}
\caption{Posterior distributions for the stellar density from isochrone fits to the spectroscopic parameters (red) and from photodynamical modeling of the lightcurve (blue). The TTV-based stellar density from L13 is also plotted with one-sigma errors for comparison (black).}
\label{fig:densities}
\end{figure}

The second major degeneracy (impact parameter, $b$) is determined by the shape of the transits. The slope of the ingress/egress indicates the curvature of the star during ingress/egress and therefore the radius of the star may be computed via $R_\star = (a/b) \cos i$, where $a$ is the semi-major axis and $i$ is the inclination. We also marginalize over these parameters, but note that the impact parameter is a positive definite quantity, and is consistent with 0 for planets d and g. Without perfectly measured transit shapes, there is some freedom to increase impact parameter away from 0 simultaneously with an increase in stellar radius so that the transit chord and thus $T_\mathrm{dur}$ is constant. If the stellar radius is decreased while the impact parameter is at or near 0, then there is no such compensatory degenerate parameter to change that would increase the transit chord, and the well-measured value of $T_\mathrm{dur}$ no longer fits the model. This results in the asymmetric photodynamically measured stellar density as shown in Fig.~\ref{fig:densities}. 

We also consider the effects of potential star spot crossing changing the apparent TTVs or transit durations. If star spots variations were contributing significantly to the fits, we would expect to see a greater reduced $\chi^2$ in transit compared to out of transit, as our transit model would not properly fit the planets' transits over star spots or faculae. This effect is not observed, strengthening our confidence in the sufficiency of our model.

\begin{table*}
\caption{Star and Planet Properties}
\label{table:den}
\centering 
\begin{tabular}{l | c c c | c c c} 
\hline
 & & $\mathcal{NSI}$ & & & $\mathcal{FSP}$ &  \\
Body & Mass (M$_\Earth$) & Radius (R$_\Earth$) & Density (g cm$^{-3}$) & Mass (M$_\Earth$) & Radius (R$_\Earth$) & Density (g cm$^{-3}$) \\
\hline
Kepler-11 b & $2.78^{+0.64}_{-0.66}$ & $1.83^{+0.07}_{-0.04} $ & $2.45^{+0.63}_{-0.62} $ & $2.83^{+0.62}_{-0.66}$ & $1.74^{+0.02}_{-0.02}$ & $ 2.96^{+0.66}_{-0.70} $  \\
Kepler-11 c & $5.00^{+1.30}_{-1.35}$ & $2.89^{+0.12}_{-0.04}$ & $1.11^{+0.32}_{-0.32}$ & $5.05^{+1.19}_{-1.37}$ & $2.75^{+0.02}_{-0.02}$ & $ 1.34^{+0.32}_{-0.36} $ \\
Kepler-11 d & $8.13^{+0.67}_{-0.66}$ & $3.21^{+0.12}_{-0.04}$ & $ 1.33^{+0.14}_{-0.15} $ & $7.52^{+0.68}_{-0.68}$ & $3.06^{+0.02}_{-0.02}$ & $ 1.45^{+0.13}_{-0.13} $ \\
Kepler-11 e & $9.48^{+0.86}_{-0.88}$ & $4.26^{+0.16}_{-0.07}$ &  $ 0.66^{+0.08}_{-0.09} $ & $8.37^{+1.01}_{-1.04}$ & $4.03^{+0.02}_{-0.03}$ & $ 0.71^{+0.09}_{-0.09} $ \\
Kepler-11 f & $2.53^{+0.49}_{-0.45}$ & $2.54^{+0.10}_{-0.04}$ & $ 0.83^{+0.18}_{-0.16} $  & $1.59^{+0.58}_{-0.54}$ & $2.40^{+0.03}_{-0.03}$ & $ 0.63^{+0.23}_{-0.21} $  \\
Kepler-11 g & $ <  27$ & $3.33^{+0.26}_{-0.09}$ & $ <  4 $ & $<  29$ & $3.16^{+0.03}_{-0.03}$ & $ < 5 $\\
\hline    
Kepler-11  & 1.04 $M_{\odot}$ (fixed) & $1.07^{+0.04}_{-0.01}$ $R_{\odot}$ & $1.19^{+0.04}_{-0.11}$  & 1.04 $M_{\odot}$ (fixed) & 1.02 $R_{\odot}$ (fixed) & 1.38 (fixed) \\
\hline   
\end{tabular}
\tablecomments{Medians and 1-$\sigma$ uncertainties from the DEMCMC runs as described in \S~\ref{s:ttvs}}

\end{table*}

\section{Discussion}
\label{s:discussion}
\subsection{Discrepancies in Stellar Densities}

The stellar densities found through spectroscopic characterization (\replaced{1.43}{1.38} $\pm$ 0.10 \gcm) and photodynamical modeling ($1.191^{+0.043}_{-0.11} $ \gcm) are inconsistent at the level of \replaced{$\geq$}{$\sim$}2$\sigma$ (Figure \ref{fig:densities}). The uncertainties on the fundamental stellar parameters would need to have been underestimated by at least a factor of 4 to allow 1-$\sigma$ agreement with the lightcurve-based stellar density measurement, which we regard as unlikely from extensive tests on our spectroscopic methods \citep{Bedell2014,Ramirez2014}. While stellar densities from fundamental parameters can be strongly dependent on imperfect stellar isochrone models, we note that in this case Kepler-11's extreme similarity to the Sun places it near the anchor point of most models, increasing the accuracy of isochronal analysis. Moreover, multiple independent age determination methods support the result of a young, non-evolved age and therefore a solar-like density for Kepler-11.

An alternative hypothesis is that some bias in the transit analysis has resulted in an erroneously low inferred stellar density. As described by \citet{Kipping2014}, multiple effects can bias the density measured by transits, including stellar activity, blended background sources, and non-zero planet eccentricities. Bias due to an underestimated planet eccentricity is not a likely explanation in this case, since all five planets give a consistent stellar density. Also, the photodynamical modeling used in this analysis should be robust to the effects of transit timing or duration variations on the measured stellar density. This leaves two potentially viable explanations from \citet{Kipping2014} for the density discrepancy: stellar activity (the ``photospot'' effect) or a background source (the ``photoblend'' effect).

Starspots effectively reduce the observed stellar flux, artificially raising stellar density inferred from the transit depth, which is the opposite of the effect we seek to explain. However, as a $\sim$3-4 Gyr Sun-like star, Kepler-11's activity may manifest mostly in the form of plages \citep{Radick1998}. Unocculted plages could potentially lower the observed stellar density by inflating the measured radii \citep{Oshagh2014}. Given the observed behavior of other main-sequence solar analogs and the lack of rotational modulation in the \Kepler lightcurve, the filling factor for spots or plages on Kepler-11's surface should be of order a few percent at most \citep{Meunier2010}. This would yield a similarly small percent-level change in the observed stellar density \citep{Kipping2014}. Furthermore, the active region configuration would need to be relatively stable throughout \textit{Kepler}'s four years of observations, which is unlikely at the high level of activity needed to have a large plage filling factor.

The final effect is blending of unresolved background sources, which can cause stellar density to be underestimated. Recently \citet{Wang2015} found two visual companions to Kepler-11 at separations of 1.36'' and 4.9'' using AO imaging. With brightness differences of $\Delta K$ = 4.4 mag and 4.7 mag respectively, these companions should contribute approximately 3\% of the total flux in the Kepler bandpass. Using Equation 9 of \citet{Kipping2014}, this implies that the observed stellar density from transits should be $\sim$99\% of the true density. The known companions are therefore insufficient to explain the magnitude of the density discrepancy.

We are left with no obvious culprit for the discrepancy between the stellar densities measured from spectroscopic characterization and lightcurve modeling. Similar testing for other systems with measured TTVs is an important next step in determining whether this is a one-off event due to, e.g. underestimated uncertainties of stellar properties or unexpected stellar activity in the lightcurve, or if it is a systematic difference between these independent methods of analysis. \replaced{If this is a systematic effect, it may be linked to the putative planet density underestimation problem in the TTV community \citep{Weiss2014}.}{If this is a systematic effect, it may be linked to the mass underestimation problem in TTV measurements relative to RVs found by \citet{Weiss2014}.}

\subsection{Implications for the Planets}
\label{s:planetprop}

The adopted mass and radius of Kepler-11 has considerable repercussions for its planetary system. We can approximate the planet mass derived from TTVs as a linear function of the assumed stellar mass. The planet radius also has a linear dependence on stellar radius, since only the relative surface areas of planet and star can be measured by the transit depth. \added{As shown in Table \ref{table:den}, fixing the stellar parameters at the spectroscopic values results in substantial changes in the planet properties which do not always follow the expected linear behavior. Because the $\mathcal{FSP}$ run did not yield a well-fitting result, as discussed in Section \ref{s:ttvs}, we do not recommend adopting these values for the planet parameters. Instead, we use the well-fitting $\mathcal{NSI}$ results and scale the planet radii to the correct stellar values by multiplying by 0.95.}

\begin{figure*}
\centering
\includegraphics[scale=0.6]{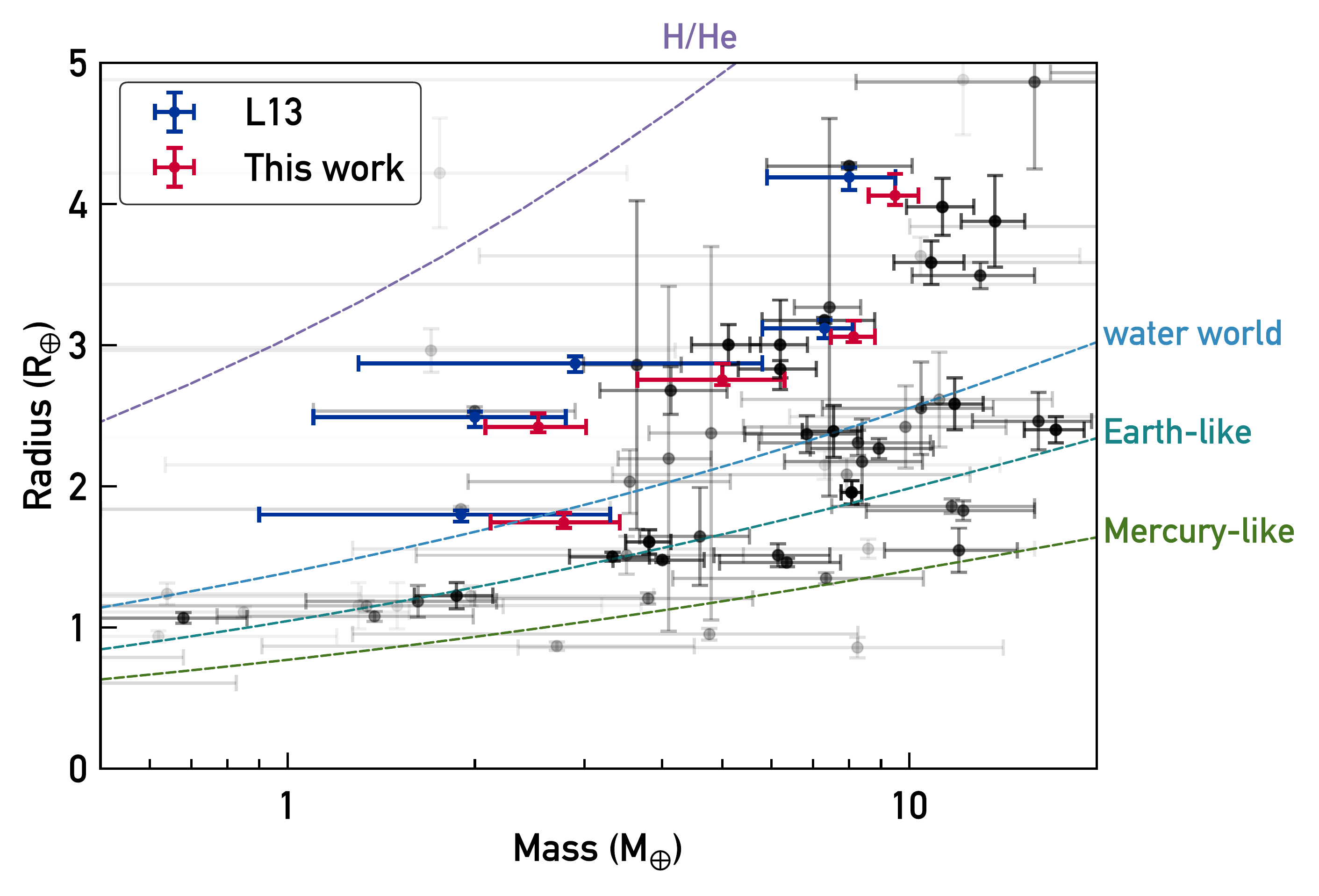}
\caption{Exoplanets with measured masses and radii. Transparencies of the black points scale with the relative error on planet parameters. Kepler-11 a-e are plotted in \replaced{red}{blue} (using the \replaced{transit and TTV-derived parameters}{L13 values}) and \replaced{blue}{red} \replaced{(adjusted by the spectroscopic stellar parameters)}{(using the spectroscopically-adjusted TTV results derived by us, as described in Section \ref{s:planetprop})}. Dashed lines show models of different compositions from \citet{Seager2007}.}
\label{fig:mr}
\end{figure*}

\replaced{The stellar properties obtained through spectroscopic analysis therefore raise the planet masses by a factor of 8\% and lower the planet radii by a factor of 5\% relative to the transit and TTV-derived values.}{Compared to previously published planet parameters from L13, our newly derived values raise the planet masses and lower the planet radii substantially, resulting in an average bulk density increase of nearly 50\%.} The results are shown in Figure \ref{fig:mr}.

\added{These parameter changes are in part a result of the new stellar mass and radius, which increased the density of the star (and thereby the inferred density of its planets) by 25\%. The remainder of the change in planet density and in particular the tightening of the constraints on the planet masses arises from the method of TTV fitting. L13 adopt very conservative error bars which stretch across the 1-$\sigma$ region of the results obtained by three independent transit time measurements. Our analysis, while less conservative, is fully self-consistent and makes use of all the available \Kepler data through a full photodynamic process. Because we fit the entire lightcurve simultaneously, our model marginalizes over any uncertainties in the transit shapes and durations, which increases our confidence in the error estimates produced. Interestingly, of the three analyses presented in L13, our results are matched most closely by the analysis which uses no a priori assumption on the transit shape and applies minimal outlier rejection to the lightcurve, much like the choices made in our own modeling.}

\replaced{Adopting the stellar properties from spectroscopic analysis raises the mean densities of the Kepler-11 planets by $\sim$30\%.}{Our analysis most substantially raises the mean densities of the two innermost planets, Kepler-11 b and c, which undergo changes of 60\% and 95\% respectively.} These increased densities, which imply a lower gas mass fraction in the planets' compositions, could make in-situ formation an increasingly viable explanation \citep[see e.g.][]{Lee2014}.

\subsection{Stellar Composition \& Planets}

While Kepler-11 is slightly more metal-rich than the Sun, its relative elemental abundances have a similar trend with \tc\ to the solar abundance pattern. Under the \citet{Melendez2009} hypothesis that the Sun's photospheric composition reflects its planet-forming history, we could interpret Kepler-11's abundance pattern as a signature of the formation of rocky planets. Such a chemical signature of terrestrial planet formation has also been revealed in Kepler-10 host star, showing the depletion of refractory materials when compared to its stellar twins \citep{Liu2016}. It is, however, somewhat dangerous to draw conclusions about the abundance pattern of an individual system, as many other factors can affect stellar abundances at the few-percent level, including galactic chemical evolution and circumstellar disk physics \citep{Gaidos2015}.

The relatively large uncertainty on the condensation temperature trend underscores the importance of galactic chemical evolution effects in particular. Although we have achieved very high-precision stellar abundance measurements, more work remains to be done on disentangling potential planet formation signatures from stellar age-dependent effects. For an individual system, even a solar twin with an age within a couple Gyr of the Sun, the uncertain effects of GCE make it extremely challenging to draw conclusions about the significance of the stellar abundance pattern in the context of planet formation. Fortunately, large-scale surveys like APOGEE and GAIA-ESO will provide the large sample sizes needed to refine \deleted{high-precision} abundance-age relations.

Regardless, it is surprising that a star that is nearly indistinguishable from the Sun even with our most advanced characterization methods is orbited by a planetary system that is so different from our own. This result continues the theme of exoplanet discoveries pointing towards a much larger variety of outcomes from the planet formation and evolution processes than was predicted even just a few years ago.

\section{Conclusion}

Using an extremely high-quality spectrum of the multi-planet host star Kepler-11, we have measured the stellar fundamental parameters and abundances to percent-level precision. We have also used a photodynamical model to fit the full \Kepler lightcurve. Our planet orbital parameters agree with past publications. However, we find that the host star is younger than previously thought by a factor of $\sim$3, with a higher \teff, \logg, and metallicity. Based on spectroscopic results, Kepler-11 and its planets are \replaced{$\sim$30\%}{20-95\%} denser than previously reported. These results stand in tension with the lightcurve results.

The five inner planets of the Kepler-11 system are key members of the exoplanet mass-radius diagram as examples of the surprisingly low densities found in some planetary systems. The substantial revision of their properties reported here underscores the importance of detailed host star follow-up. As the community looks to exponentially increase the number of exoplanets with measured bulk densities through TESS and beyond, it is critical to prioritize securing high-quality spectra of the host stars to enable the determination of precise host star properties.

\bigskip
\acknowledgements

\added{We thank John Southworth, Pierre Maxted, and the anonymous referee for their helpful comments on the first draft of this paper.}

This work was supported by a NASA Keck PI Data Award, administered by the NASA Exoplanet Science Institute. Data presented herein were obtained at the W. M. Keck Observatory from telescope time allocated to the National Aeronautics and Space Administration through the agency's scientific partnership with the California Institute of Technology and the University of California. The Observatory was made possible by the generous financial support of the W. M. Keck Foundation.

The authors wish to recognize and acknowledge the very significant cultural role and reverence that the summit of Mauna Kea has always had within the indigenous Hawaiian community. We are most fortunate to have the opportunity to conduct observations from this mountain.

M.B. is supported by a National Science Foundation Graduate Research Fellowship under Grant No. DGE-1144082.  J.L.B. acknowledges support for this work from the NSF (grant number AST-1313119) and the Packard Foundation.  J.M. thanks FAPESP (2012/24392-2). M.A., D.Y., and F.L. have been supported by the Australian Research Council
(grants FL110100012, DP120100991 and FT140100554).

\facilities{Keck:I (HIRES), Kepler}


\bibliographystyle{apj}
\bibliography{}

\pagebreak

\begin{longrotatetable}
\begin{deluxetable*}{cccccccccccccccc} 
\centering
\tablecaption{Line List and Measured Equivalent Widths.\label{tbl:ews}}
\tabletypesize{\scriptsize}
\tablehead{
\colhead{Wavelength} & \colhead{Species\tablenotemark{1}} & 
\colhead{EP} & \colhead{log($gf$)} & \colhead{Kepler-11} & \colhead{Sun} & \colhead{Sun2\tablenotemark{2}} &
\colhead{HD1178} & \colhead{HD10145} & 
\colhead{HD16623} & \colhead{HD20329} & 
\colhead{HD21727} & \colhead{HD21774} & \colhead{HD28474} &
\colhead{HD176733} & \colhead{HD191069} \\
\colhead{(\r{A})} & \colhead{} & \colhead{eV} & \colhead{(dex)} &
\colhead{(m\r{A})} & \colhead{(m\r{A})} & \colhead{(m\r{A})} & 
\colhead{(m\r{A})} & \colhead{(m\r{A})} & \colhead{(m\r{A})} & 
\colhead{(m\r{A})} & \colhead{(m\r{A})} & \colhead{(m\r{A})} & 
\colhead{(m\r{A})} & \colhead{(m\r{A})} & \colhead{(m\r{A})}
} 
\startdata
5052.167 & 6.0  & 7.685 & $-$1.24 & 35.2 & 33.5 & & 32.7 & 27.9 & 21.4 & 29.4 & 26.6 & 41.6 & 17 & 27.4 & 36.7 \\
6587.61 & 6.0 & 8.537 & $-$1.05 & 16.9 & 15.2 & & 17 & 12.2 & 9.8 & 11.4 & 11.7 & 20.3 & 6.2 & 10.8 & 17 \\
7111.47 & 6.0 & 8.64 & $-$1.07 & 13.2 & 11.3 & & 12.1 & 15.1 & 9.6 & 12.3 & 10.9 & 16.6 & 36.7 & 10.2 & 14 \\
7113.179 &  6.0 & 8.647 & $-$0.76 & 25.3 & 22.4 & & 23.2 & 20.3 & 13.8 & 17.9 & 19.5 & 33 & 11.3 & 18.2 & 24.3 \\
7771.944 & 8.0 & 9.146 & 0.37 & 77.8 & 69.7 & & 79.1 & 63.6 & 71 & 69.4 & 62.9 & 79.9 & 56.9 & 61.5 & 81.8 \\
  &  & & & & & & $\vdots$   & &  & & & & & \\
\enddata
\tablenotetext{1}{Where the number before the decimal is the element number and the number after the decimal is the ionization state, e.g. 6.0 = C I.}
\added{\tablenotetext{2}{Solar equivalent widths used as reference for HD10145, HD21727, and HD191069 stellar parameters. The remaining stars use the other solar measurements.}}
\tablecomments{Table \ref{tbl:ews} is published in its entirety in the machine-readable format. A portion is shown here for guidance regarding its form and content.}
\end{deluxetable*}
\end{longrotatetable}

\pagebreak

\begin{longrotatetable}
\begin{deluxetable*}{lcccccccccccc} 
\tablecaption{Differential Abundances [X/H].\label{tbl:abund}}
\tablewidth{900pt}
\tabletypesize{\scriptsize}
\tablehead{
\colhead{Element} & \colhead{$T_{c}$\tablenotemark{1}} & 
\colhead{$N_{lines}$} & \colhead{Kepler-11} & 
\colhead{HD1178} & \colhead{HD10145} & 
\colhead{HD16623} & \colhead{HD20329} & 
\colhead{HD21727} & \colhead{HD21774} & \colhead{HD28474} &
\colhead{HD176733} & \colhead{HD191069}
} 
\startdata
CI & 40 & 4 & 0.027 $\pm$ 0.011 & 0.122 $\pm$ 0.022 & 0.059 $\pm$ 0.069 & $-$0.204 $\pm$ 0.053 & 0.053 $\pm$ 0.046 & 0.024 $\pm$ 0.037 & 0.182 $\pm$ 0.034 & $-$0.090 $\pm$ 0.265 & 0.007 $\pm$ 0.030 & 0.084 $\pm$ 0.017 \\
OI & 180 & 3 & 0.043 $\pm$ 0.011 & 0.192 $\pm$ 0.055 & 0.073 $\pm$ 0.021 & 0.004 $\pm$ 0.031 & 0.212 $\pm$ 0.025 & 0.079 $\pm$ 0.022 & 0.150 $\pm$ 0.034 & $-$0.159 $\pm$ 0.029 & 0.059 $\pm$ 0.022 & 0.167 $\pm$ 0.017 \\
NaI & 958 & 4 & 0.045 $\pm$ 0.010 & $-$0.027 $\pm$ 0.012 & $-$0.119 $\pm$ 0.028 & $-$0.389 $\pm$ 0.017 & $-$0.035 $\pm$ 0.022 & $-$0.088 $\pm$ 0.014 & 0.267 $\pm$ 0.024 & $-$0.558 $\pm$ 0.032 & $-$0.027 $\pm$ 0.015 & $-$0.020 $\pm$ 0.013 \\
MgI & 1336 & 5 & 0.035 $\pm$ 0.014 & 0.098 $\pm$ 0.017 & 0.040 $\pm$ 0.010 & $-$0.228 $\pm$ 0.031 & 0.047 $\pm$ 0.059 & 0.042 $\pm$ 0.032 & 0.258 $\pm$ 0.032 & $-$0.411 $\pm$ 0.033 & 0.025 $\pm$ 0.022 & 0.078 $\pm$ 0.011 \\
AlI & 1653 & 2 & 0.061 $\pm$ 0.008 & 0.176 $\pm$ 0.018 & 0.033 $\pm$ 0.007 & $-$0.297 $\pm$ 0.018 & 0.177 $\pm$ 0.025 & 0.065 $\pm$ 0.007 & 0.279 $\pm$ 0.018 & $-$0.469 $\pm$ 0.009 & 0.042 $\pm$ 0.006 & 0.105 $\pm$ 0.011 \\
SiI & 1310 & 14 & 0.047 $\pm$ 0.005 & 0.038 $\pm$ 0.009 & $-$0.009 $\pm$ 0.004 & $-$0.285 $\pm$ 0.011 & 0.042 $\pm$ 0.013 & 0.012 $\pm$ 0.008 & 0.245 $\pm$ 0.009 & $-$0.461 $\pm$ 0.011 & $-$0.004 $\pm$ 0.006 & 0.033 $\pm$ 0.004 \\
SI & 664 & 4 & 0.036 $\pm$ 0.023 & 0.106 $\pm$ 0.022 & 0.046 $\pm$ 0.019 & $-$0.249 $\pm$ 0.042 & 0.034 $\pm$ 0.050 & 0.015 $\pm$ 0.016 & 0.235 $\pm$ 0.026 & $-$0.385 $\pm$ 0.055 & 0.031 $\pm$ 0.032 & 0.094 $\pm$ 0.030 \\
KI & 1006 & 1 & 0.068 $\pm$ 0.013 & 0.031 $\pm$ 0.012 & $-$0.037 $\pm$ 0.017 & $-$0.168 $\pm$ 0.038 & $-$0.025 $\pm$ 0.011 & $-$0.079 $\pm$ 0.015 & 0.192 $\pm$ 0.040 & $-$0.355 $\pm$ 0.028 & $-$0.077 $\pm$ 0.016 & 0.045 $\pm$ 0.011 \\
CaI & 1517 & 11 & 0.066 $\pm$ 0.009 & 0.072 $\pm$ 0.011 & 0.008 $\pm$ 0.011 & $-$0.301 $\pm$ 0.022 & 0.052 $\pm$ 0.015 & 0.025 $\pm$ 0.013 & 0.230 $\pm$ 0.028 & $-$0.494 $\pm$ 0.014 & $-$0.003 $\pm$ 0.010 & 0.014 $\pm$ 0.007 \\
ScI & 1659 & 4 & 0.086 $\pm$ 0.022 & 0.096 $\pm$ 0.019 & 0.021 $\pm$ 0.018 & $-$0.335 $\pm$ 0.042 & 0.067 $\pm$ 0.036 & 0.024 $\pm$ 0.017 & 0.271 $\pm$ 0.032 & $-$0.368 $\pm$ 0.047 & 0.005 $\pm$ 0.026 & 0.051 $\pm$ 0.010 \\
ScII & 1659 & 5 & 0.101 $\pm$ 0.013 & 0.140 $\pm$ 0.028 & $-$0.003 $\pm$ 0.013 & $-$0.288 $\pm$ 0.032 & 0.144 $\pm$ 0.027 & 0.077 $\pm$ 0.025 & 0.316 $\pm$ 0.031 & $-$0.465 $\pm$ 0.031 & 0.004 $\pm$ 0.011 & 0.091 $\pm$ 0.022 \\
TiI & 1582 & 18 & 0.065 $\pm$ 0.009 & 0.113 $\pm$ 0.010 & 0.029 $\pm$ 0.015 & $-$0.244 $\pm$ 0.026 & 0.148 $\pm$ 0.011 & 0.041 $\pm$ 0.011 & 0.243 $\pm$ 0.033 & $-$0.451 $\pm$ 0.016 & 0.009 $\pm$ 0.011 & 0.053 $\pm$ 0.008 \\
TiII & 1582 & 11 & 0.070 $\pm$ 0.013 & 0.108 $\pm$ 0.017 & $-$0.024 $\pm$ 0.032 & $-$0.229 $\pm$ 0.032 & 0.117 $\pm$ 0.013 & $-$0.004 $\pm$ 0.033 & 0.255 $\pm$ 0.034 & $-$0.423 $\pm$ 0.025 & $-$0.007 $\pm$ 0.013 & 0.076 $\pm$ 0.014 \\
VI & 1429 & 9 & 0.078 $\pm$ 0.009 & 0.078 $\pm$ 0.010 & $-$0.014 $\pm$ 0.014 & $-$0.323 $\pm$ 0.029 & 0.097 $\pm$ 0.032 & 0.027 $\pm$ 0.014 & 0.275 $\pm$ 0.033 & $-$0.530 $\pm$ 0.022 & 0.009 $\pm$ 0.012 & 0.011 $\pm$ 0.009 \\
CrI & 1296 & 10 & 0.047 $\pm$ 0.009 & 0.017 $\pm$ 0.009 & $-$0.006 $\pm$ 0.020 & $-$0.484 $\pm$ 0.023 & $-$0.065 $\pm$ 0.011 & $-$0.008 $\pm$ 0.009 & 0.272 $\pm$ 0.029 & $-$0.626 $\pm$ 0.021 & 0.001 $\pm$ 0.010 & $-$0.042 $\pm$ 0.008 \\
CrII & 1296 & 5 & 0.055 $\pm$ 0.013 & 0.010 $\pm$ 0.013 & $-$0.047 $\pm$ 0.028 & $-$0.415 $\pm$ 0.035 & $-$0.073 $\pm$ 0.011 & $-$0.010 $\pm$ 0.018 & 0.243 $\pm$ 0.032 & $-$0.586 $\pm$ 0.033 & $-$0.025 $\pm$ 0.015 & $-$0.044 $\pm$ 0.012 \\
MnI & 1158 & 8 & 0.060 $\pm$ 0.011 & $-$0.048 $\pm$ 0.011 & $-$0.084 $\pm$ 0.013 & $-$0.634 $\pm$ 0.022 & $-$0.183 $\pm$ 0.009 & $-$0.065 $\pm$ 0.010 & 0.298 $\pm$ 0.028 & $-$0.778 $\pm$ 0.025 & $-$0.028 $\pm$ 0.011 & $-$0.105 $\pm$ 0.008 \\
FeI & 1334 & 92 & 0.062 $\pm$ 0.010 & 0.013 $\pm$ 0.010 & $-$0.033 $\pm$ 0.013 & $-$0.462 $\pm$ 0.023 & $-$0.093 $\pm$ 0.010 & $-$0.018 $\pm$ 0.013 & 0.252 $\pm$ 0.030 & $-$0.613 $\pm$ 0.014 & $-$0.018 $\pm$ 0.011 & $-$0.057 $\pm$ 0.013 \\
FeII & 1334 & 16 & 0.064 $\pm$ 0.010 & 0.022 $\pm$ 0.015 & $-$0.031 $\pm$ 0.019 & $-$0.445 $\pm$ 0.028 & $-$0.085 $\pm$ 0.014 & $-$0.021 $\pm$ 0.018 & 0.260 $\pm$ 0.034 & $-$0.601 $\pm$ 0.024 & $-$0.019 $\pm$ 0.024 & $-$0.043 $\pm$ 0.015 \\
CoI & 1352 & 6 & 0.062 $\pm$ 0.044 & 0.053 $\pm$ 0.044 & $-$0.035 $\pm$ 0.039 & $-$0.314 $\pm$ 0.075 & 0.023 $\pm$ 0.039 & $-$0.006 $\pm$ 0.050 & 0.260 $\pm$ 0.053 & $-$0.499 $\pm$ 0.072 & $-$0.004 $\pm$ 0.073 & 0.032 $\pm$ 0.061 \\
NiI & 1353 & 20 & 0.066 $\pm$ 0.007 & 0.010 $\pm$ 0.007 & $-$0.052 $\pm$ 0.008 & $-$0.443 $\pm$ 0.018 & $-$0.061 $\pm$ 0.009 & $-$0.022 $\pm$ 0.010 & 0.278 $\pm$ 0.022 & $-$0.631 $\pm$ 0.012 & $-$0.012 $\pm$ 0.010 & $-$0.030 $\pm$ 0.006 \\
CuI & 1037 & 4 & 0.080 $\pm$ 0.008 & 0.075 $\pm$ 0.019 & $-$0.037 $\pm$ 0.009 & $-$0.472 $\pm$ 0.036 & 0.000 $\pm$ 0.008 & $-$0.021 $\pm$ 0.013 & 0.325 $\pm$ 0.028 & $-$0.630 $\pm$ 0.053 & 0.053 $\pm$ 0.020 & 0.034 $\pm$ 0.012 \\
ZnI & 726 & 3 & 0.036 $\pm$ 0.028 & 0.036 $\pm$ 0.012 & $-$0.020 $\pm$ 0.017 & $-$0.315 $\pm$ 0.037 & 0.056 $\pm$ 0.023 & $-$0.034 $\pm$ 0.007 & 0.283 $\pm$ 0.043 & $-$0.539 $\pm$ 0.062 & $-$0.007 $\pm$ 0.009 & 0.070 $\pm$ 0.016 \\
YII & 1659 & 5 & 0.078 $\pm$ 0.017 & $-$0.025 $\pm$ 0.012 & $-$0.109 $\pm$ 0.017 & $-$0.579 $\pm$ 0.033 & $-$0.123 $\pm$ 0.018 & $-$0.088 $\pm$ 0.013 & 0.228 $\pm$ 0.041 & $-$0.673 $\pm$ 0.030 & $-$0.051 $\pm$ 0.017 & $-$0.124 $\pm$ 0.011 \\
CH & 40 & 3 & 0.054 $\pm$ 0.007 & $-$0.006 $\pm$ 0.011 & $-$0.088 $\pm$ 0.015 & $-$0.436 $\pm$ 0.028 & $-$0.199 $\pm$ 0.014 & $-$0.122 $\pm$ 0.018 & 0.225 $\pm$ 0.027 & $-$0.620 $\pm$ 0.019 & $-$0.031 $\pm$ 0.019 & $-$0.028 $\pm$ 0.019 \\
\enddata
\tablenotetext{1}{50\% condensation temperature for the element under protoplanetary disk conditions from \citet{Lodders2003}.}
\end{deluxetable*}
\end{longrotatetable}

\end{document}